# A Comparative Evaluation of Statistical Methods in Hybrid Controlled Trials

Di Ran[a]*, Fanni Zhang[a], Sima Shahsavari[b], Kristine Broglio[a], Alasdair Henderson[c], and Binbing Yu[a]

[a]*Oncology Biometrics Statistical Innovation, AstraZeneca, Gaithersburg, USA;* [b]*Late CVRM Biometrics, AstraZeneca, City, Gothenburg, Sweden;* [c]*School of Cardiovascular and Metabolic Health, University of Glasgow, UK*

*Corresponding author: Di Ran, di.ran@astrazeneca.com, Oncology Biometrics Statistical Innovation, AstraZeneca, 1 MedImmune Way, Gaithersburg, Maryland, 20878, USA

# A Comparative Evaluation of Statistical Methods in Hybrid Controlled Trials

Randomized clinical trials (RCTs) are widely considered the gold standard for evaluating the effectiveness of new treatments or interventions in drug development. Still, they may not be feasible in certain cases, such as with rare diseases where randomization to a control group is ethically challenging. In such scenarios, external data can complement either a single-arm trial or a hybrid-controlled trial. The hybrid-control design involves enrolling fewer concurrent control patients and then supplementing the control arm using external or historical data. Various statistical approaches, including frequentist methods (e.g., propensity score methods), Bayesian borrowing approaches (e.g., meta-analytic-predictive prior), and their integration, have been utilized to incorporate external information in hybrid-controlled trials. We evaluate several accessible methods for their robustness to between-study heterogeneity and unmeasured confounding utilizing a case study based on data from the DAPA-HF trial, along with comprehensive simulation studies. Our findings indicate that the optimal methods must take into account the heterogeneities from both measured and unmeasured confounding. Since no single method consistently outperforms all others, researchers should explore multiple methods through extensive simulations to evaluate their effectiveness across various scenarios.

Keywords: Hybrid-controlled trial; external data; propensity score; Bayesian borrowing

**Introduction**

Randomized controlled clinical trials (RCTs) have long been regarded as the "gold standard" for drug development (Food and Drug Administration 1998; Peck et al. 2003). However, RCTs often face challenges such as prolonged timelines, high costs, and difficulties in enrolling patients, especially when dealing with rare medical conditions (Food and Drug Administration 2019). In response, regulatory interest in leveraging external data sources into clinical trials has increased, including historical clinical trial data and real-world data (RWD) (Food and Drug Administration 2021, 2023). Notably, external data can enrich our understanding of standard-of-care control arms (Food and Drug Administration 2001). Innovative trial design and analyses are necessary to incorporate such external data into RCTs. These approaches aim to improve the efficiency of efficacy and safety assessments while retaining the robustness of RCT conclusions (Neuenschwander et al. 2010; Hernán and Robins 2016).

Single-arm trials may be one choice where external data for a standard-of-care control arm can function as a formal comparator arm, also referred to as a synthetic control arm, to establish the effectiveness of a new therapy, especially for rare diseases (Thorlund et al. 2020). An alternative is the hybrid control arm design, which integrates external data to augment a concurrent control arm of an RCT (Pocock 1976; Food and Drug Administration 2023). In such a design, the randomization ratio in the concurrent trial often favors the investigational drug (e.g., 2:1 or 3:1 between treated and control groups), making the trial more attractive to patients. Augmenting the smaller concurrent control arm with external data reduces the total sample size required, which can accelerate the trial timelines. This paper focuses on methods for incorporating external data in this hybrid control arm setting. We use the term "historical trials" throughout to refer to the set of external data sources that are being incorporated, where only subjects that received the same control arm therapy would be selected for inclusion.

When considering external controls in clinical research, it is first essential to assess whether the data source is fit for purpose. This involves ensuring the data can reliably address the scientific questions and meet requirements to support regulatory decision-making (Levenson et al. 2023). Food and Drug Administration (FDA) guidance documents highlight regulatory concerns with the use of external control arms, such as selection bias, unmeasured confounding, lack of concurrency, and the validity of covariates and outcomes (Food and Drug Administration 2019, 2023). Ignoring these concerns may bias the assessment of effectiveness and could result in misleading conclusions. Selection bias often arises when subjects in historical trials have different characteristics from the subjects in the trial. To mitigate these differences, statistical methods have been proposed to efficiently borrow information from historical trials while maintaining reliable inferences.

We evaluate common and easily accessible methods for augmenting a control arm in a hybrid clinical trial for their robustness to between-study heterogeneity and unmeasured confounders. We provide a brief review of the selected methods, which include propensity score (PS) methods, Bayesian information borrowing methods, and PS-integrated approaches. We first apply the methods in a case study, using the data from the DAPA-HF randomization clinical trial (McMurray et al. 2019). We then evaluate these methods via a comprehensive simulation study, covering a wide range of scenarios varying the number of external data sources, the between-study heterogeneities, and the degree of unmeasured confounding.

**Methods and Materials**

*Review of statistical methods for hybrid-controlled trials*

Both frequentist and Bayesian approaches can be used to borrow information from historical trials. A key difference among these methods is the way that they mitigate the confounding. Some of them, like PS methods, incorporate the measured confounding directly, with a strong assumption of no unmeasured confounding. In contrast, many Bayesian approaches address heterogeneities across the study summary level outcome measures, perhaps ignoring the existing information of the measured confounders. Other methods, such as mixed-effect models and PS-integrated methods, take advantage of both perspectives.

Propensity score (PS) methods, a set of frequentist approaches, are often used for nonrandomized studies to address the imbalance in the distribution of observed baseline covariates (Rosenbaum and Rubin 1983). There are four common propensity score methods: matching, stratification, inverse probability of treatment weighting, and covariate adjustment using the propensity score (Austin 2011), with the first three being more popular (Austin 2010). Propensity score matching, for example, has been demonstrated to enhance a control arm in an RCT, as evidenced in case studies (Lin et al. 2018). The PS methods help ensure that the outcome data remain blinded during the design stage, which is beneficial in a regulatory setting (Lin et al. 2018). Other frequentist approaches, like mixed-effect models, can borrow information from historical trials while accounting for the between-trial heterogeneities (Wu 2009). The alternatives to the frequentist approaches are Bayesian approaches, including the Bayesian hierarchical model (Neuenschwander et al. 2010), power prior (PP) (Ibrahim and Chen 2000), commensurate prior (Hobbs et al. 2012), meta-analytic predictive prior

(Neuenschwander et al. 2010), and varieties based on these methods, such as the robust meta-analytic predictive prior (MAP) (Schmidli et al. 2014). Bayesian methods use a strategy of dynamically weighting information from historical trials according to their similarity and dissimilarity to the concurrent trial, without relying on strong assumptions such as the absence of unmeasured confounding in PS methods. Simulation studies evaluating the operating characteristics of many of these Bayesian approaches in the hybrid trial setting have shown that the meta-analytic predictive prior is the most promising method, particularly when multiple historical trials are available (van Rosmalen et al. 2018; Su et al. 2022).

More recently, novel methods have been developed to integrate propensity score methods with other frequentist or Bayesian techniques, allowing them to inherit the advantages of both PS and Bayesian methods. Several of these methods are built-upon propensity score stratification, including propensity score stratification integrated with either the power prior (Wang et al. 2019; Lu et al. 2022), composite likelihood (Chen et al. 2020), or meta-analytic predictive prior (Liu et al. 2021). In general, these methods stratify patients based on their propensity scores, apply frequentist or Bayesian borrowing methods within each stratum, and then summarize the results across strata. Another way to take advantage of both PS and Bayesian approaches is to simply use a PS-matched or weighted sample for the meta-analytic predictive prior, thus combining the most popular PS methods with the most promising Bayesian information borrowing method. We identified six methods that have publicly available implementations in R by conducting a literature review. We also note that Bayesian methods, such as the Bayesian hierarchical model, power prior, and commensurate prior, along with several variants of these methods, have been thoroughly examined in the previous simulation studies, and the MAP prior approach is the most promising one, especially when multiple external studies are available (van Rosmalen et al. 2018; Su et al. 2022). Thus, we choose MAP as the representative Bayesian approach. Propensity score stratification integrated with MAP is not included in the assessment because no ready-to-use R implementation is available. However, we add two easy-to-implement integrated approaches, incorporating PS matching or weighting with the MAP approach.

All these statistical methods can account for the heterogeneities among trials but mitigate the potential bias via different techniques. In this section, we provide a brief description of each. We summarize their unique features, requirements, and R implementations in **Table 1**. Note that, although we summarize the PSS+CL approach in this method section and the summary table, it is excluded from the case study and simulation results in the main text due to the unsound SE estimates using the R package. The results of PSS+CL can be found in the supplementary materials.

*Propensity score methods*

Propensity score methods, widely employed to mitigate confounding bias in nonrandomized data, can be used in hybrid-controlled trials, where the propensity score is defined as the probability that a subject is enrolled in the concurrent trial ($T_i = 1$ if $i \in \{c\}$ vs. historical trial, $T_i = 0$, if $i \in \{H_1, H_2, \ldots, H_k\}$) conditional on a set of observed covariates ($X$): $e_i(X_i) = \Pr(T_i = 1|X_i)$, $i$ is the $i$th subject in the pooled data. The propensity scores for individual subjects are estimated via a logistic regression model with $T$ as the outcome and $X$ as the predictors. Based on the similarity of propensity scores, PS matching pairs subjects in the concurrent trial with subjects from the historical trials. To accommodate the potential large heterogeneity among these trials, we opt for 1:1 matching with replacement and employ a caliper of 0.2. Adding these matched external controls to the concurrent trial, we recover the concurrent trial to a 1:1 "randomized" trial. Inverse probability weighting using propensity scores is another way to utilize the same propensity scores to address the heterogeneity among trials. It creates a synthetic sample that represents the concurrent trial population, where the weights are $w_i = 1$ for subjects in the concurrent trial and $w_i = \frac{e_i(X_i)}{1-e_i(X_i)}$ for historical controls. To reduce the variance of weighted estimates and bias from the confounding in the tails of the PS distribution, we employ a symmetric trimming that removes subjects with extremely small weights less than 0.05 or extremely large weights greater than 20 (Stürmer et al. 2021).

*Meta-analytic-predictive prior*

The MAP prior synthesizes the evidence from historical trials to create a predictive distribution for the parameter of interest in the concurrent trial ($\theta_c$). Let assume the concurrent data $Y_c$ can be described by a statistical model $p(Y_c|\theta_c)$. The MAP prior utilizes a Bayesian hierarchical model across historical trials to derive the predictive posterior distribution of the parameter of interest (Neuenschwander et al. 2010). The posterior for the parameter of interest in the new trial $\theta_c$ is $p(\theta_c|Y_c) \propto p(Y_c|\theta_c)p_{MAP}(\theta_c)$, where the marginal posterior for the parameter of interest, $p_{MAP}(\theta_c) = p(\theta_c|Y_{H_1}, Y_{H_2}, \dots, Y_{H_k})$, represents the prior information from the $k$ historical trials, i.e., the MAP prior. This approach is illustrated in Supplementary Figure 2a. The MAP prior can be derived using the estimated study-specific mean $\hat{\theta}_{H_j}$ and standard error $\hat{S}^2_{H_j}, j = 1, 2, \dots, k$. Further, this predictive prior can be robustified by combining a weakly-informative mixture component, i.e., $p_{rMAP}(\theta_c) \sim (1 - \omega) *$

$p_{MAP}(\theta_c) + \omega * p_V(\theta_c)$, where $p_V$ represents the weakly-informative prior and $\omega$ is the weight of this component; a weight of $\omega = 0$ implies full reliance on the historical data, while $\omega = 1$ assigns full weight to the non-informative prior. In the MAP prior, the study-specific mean is often assumed to follow a normal distribution $\theta_{H_j}|\mu, \tau^2 \sim N(\mu, \tau^2), j = 1, 2, \ldots, k$, with an overall effect size $\mu$ and between-trial variance, $\tau^2$. So, in addition to the weight parameter, the choice of a hyperparameter $\tau$, describing the between-trial heterogeneities in the prior distribution can also determine the strength of borrowing. The optimal parameter of $\omega$ and/or $\tau$ needs to be determined.

*Mixed-effects model*

Another analysis method that can accommodate heterogeneity between historical trials is the frequentist mixed-effects model (MM). We use a mixed-effects model with a random intercept, which accounts for unmeasured sources of heterogeneity between trials, $y = Z\theta + \boldsymbol{X\beta} + D\gamma + e$, where $Z$ is the treatment indicator ($Z = 1$ for treated subjects and $Z = 0$ for control subjects), $\theta$ is the parameter of interest, $D$ is the trial index ($D = 1$ if the subject in the concurrent study; otherwise, $D = 2$ corresponds to the 1st historical trial, increasing sequentially up to $D = k + 1$ for the $k$th historical trial), $\boldsymbol{\beta}$ are the fixed effects regression coefficients, $\gamma$ represents the random effect, which explains additional unmeasured systematic differences among trials. We acknowledge that a more complex model, such as one including random slopes, could potentially capture additional heterogeneities. However, exploring such complexity falls beyond the scope of this comparative study. Furthermore, the case study demonstrates that utilizing a simpler model is adequate for our research.

*PS stratification integrated composite likelihood or power prior*

Both the PS-integrated composite likelihood (PSS+CL) (Chen et al. 2020) and PS-integrated power prior (PSS+PP) (Wang et al. 2019; Lu et al. 2022) approaches build upon PS stratification to mitigate known confounding effects. Subjects in the concurrent trial and historical controls are pooled and divided into multiple strata based on their propensity scores. Within each stratum, the subjects are assumed to have similar distributions of covariates and therefore homogenous treatment effects. Two methods adopt different approaches to borrow information from historical controls and estimate the stratum-specific treatment effects. The PSS+CL utilizes a composite likelihood function, whereas the PSS+PP applies a power prior approach to appropriately weight the historical controls. Both methods then aggregate these stratum-specific estimates to obtain an overall population-level treatment effect estimate. More details can be found in the Supplementary Materials.

*PS matching/weighting integrated with MAP*

In the previous two integrated methods, the advantages come from utilizing PS methods to address measurable sources of confounding and further managing unmeasured between-trial heterogeneities using borrowing methods. Based on that principle, we consider integrating PS methods with MAP in a two-step manner, firstly creating an (observed) covariates balanced sample via PS matching or weighting, and secondly applying MAP to this sample. For propensity score matching followed by MAP (PSM+MAP), a propensity score matching with replacement is first done to identify subjects with similar covariate distributions in the historical trials compared to the subjects in the concurrent study. Next, MAP is conducted with only these PS-matched subjects to further control the between-trial heterogeneities that are not explained by the observed covariates. Specifically, the study-specific mean $\hat{\theta}_{H_j,M}$ and standard errors

$\hat{S}^2_{H_j,M}$ are estimated on the matched subjects for each historical trial $j = 1, 2, \ldots, k$ and used for the MAP prior (Supplementary Figure 2b). Similarly, for propensity score weighting followed by MAP (PSW+MAP), we first compute the inverse probability weights of propensity scores for the subjects in the historical trials as described for PSW above. Next, the weighted mean $\hat{\theta}_{H_j,W}$ and standard error $\hat{S}^2_{H_j,W}$ are estimated for each specific historical trial $j = 1, 2, \ldots, k$. As illustrated in Supplementary Figure 2c, these weighted estimates are then used to build the MAP prior. We include the exploration of the key MAP parameters in both the case study and the simulation studies.

### *Case study based on DAPA-HF data*

To demonstrate the application of the methods, we utilized a case study based on patient-level data from the DAPA-HF clinical trial, which assessed the safety and efficacy of dapagliflozin in reducing the risk of worsening heart failure or cardiovascular death in patients with heart failure, irrespective of their diabetes status (McMurray et al. 2019). Our analytical focus for this case study is on the change from baseline to 8 months in the log-transformed N-terminal pro-B-type natriuretic peptide (NT-proBNP) levels. NT-proBNP is secreted by cardiac myocytes in response to increased cardiac wall stress, a key indicator of heart failure, making it a critical diagnostic and prognostic biomarker. We limited our analysis to patient-level data from the placebo arm, with data available for 1,892 patients after excluding those with missing outcomes, incomplete baseline characteristics (encompassing eighteen variables), and those who withdrew consent due to personal reasons or data governance policies in specific countries (Broglio et al. 2025).

To evaluate the statistical methods, we divided the data into one "concurrent" study and one "historical" study based on three prognostic variables, including the baseline NT-proBNP, which is not strongly correlated with the outcome ($\rho = -0.22$) and two additional synthetic variables that were created at the patient level to have a prespecified correlation with outcome ($\rho = 0.5$ and $0.4$) (Broglio et al. 2025). Synthetic variables were used because no other baseline characteristics were identified as strong prognostic factors for the outcome in this dataset. Thus, the data sources vary in terms of 3 key patient characteristics. These three variables are related to the differences in outcome and can explain study heterogeneity. The treatment label was randomly assigned to each subject in the concurrent and historical studies. If study heterogeneity is due to only measured confounders, and the analysis recognizes this, we would expect that borrowing information from the historical trial can augment the power of the concurrent trial without introducing any bias. As such, two model specifications for the PS model and mixed-effect model were examined: a correct model comprising all three confounders and an incorrect model that omitted the two synthetic covariates to purposely create unmeasured confounding between the data sources. Because only control arm patients are used, the null hypothesis that there is no treatment effect is true, and we can assess Type I error rates only. The details of data generation can be found in the supplemental materials (supplementary Figures 3-6).

### *Simulation studies*

*Data generation*

Simulation studies were conducted to first mimic the case study as well as further consider scenarios that could not be explored in the case study. Full details of the data simulation can be found in the appendix. In brief, we use a normally distributed endpoint and assume 6 independent covariates. Coefficients within a linear regression are selected for the covariates, and the resulting patient-level outcomes are used to assign patients to a data source, i.e., the concurrent trial or a historical data source. This creates an association between patient characteristics and data source. A second set of coefficients is used to then generate the endpoint of interest at the patient level according to the covariates. Thus, patient characteristics vary across the data sources, but the differences are all explainable by the full set of 6 covariates. If the data is analysed with the full set of covariates, the assumption of no unmeasured confounding is satisfied. We refer to this as the correct model. We also consider incorrect models, where one or more covariates are omitted from the analysis. This creates an analysis setting with unmeasured confounding. The selection of the coefficients can control the degree of heterogeneity between data sources and the degree of unmeasured confounding can be controlled by the number of covariates omitted from the model. Simulation scenarios include, 1) one versus multiple historical studies available, 2) the heterogeneity across studies varied from moderate to severe, 3) the treatment effect was either null or at an alternative hypothesis treatment effect, and 4) model specifications for analysis included the correct model and two incorrect models where one or three covariates was omitted. Each simulated dataset had a total of 1,600 subjects with approximately 25% (i.e., 400) allocated to the concurrent trial and the remainder distributed among the one or three historical controls. The data generation process was replicated $M = 2000$ times for each scenario and setting to ensure reliability and statistical validity in our findings.

*Performance assessment*

For each method, we estimated the mean treatment effect along with the average standard error (SE). We also evaluated performance based on several criteria: (1) the average bias of the mean treatment effect, where $bias = \frac{1}{M}\sum_{m=1}^{M}(\hat{\theta}_{treat} - \theta_{treat})$; (2) the Type I error, defined as the probability that the null hypothesis is rejected when the null hypothesis is true; or conversely (3) the statistical power, defined as the percentage of times the null hypothesis is correctly rejected when the alternative hypothesis is true; (4) effective sample size rate (ESSR), calculated as the percentage of patients added into the current study that reduces the variation the same amount as borrowing,

$$ESSR = \left(\frac{precision(\hat{\theta}_{treat\_borrow})}{precision(\hat{\theta}_{treat\_no\_borrow})} - 1\right) \times 100\% = \left(\frac{\text{var}(\hat{\theta}_{treat\_no\_borrow})}{\text{var}(\hat{\theta}_{treat\_brrow})} - 1\right) \times 100\%.$$

ESSR is derived based on the definition of effective sample size in (Hobbs et al. 2013) and is an indication of the degree of borrowing.

**Results**

*Case study based on DAPA-HF*

Results across all the sets of considered methods based on the DAPA-HF trial data are shown in **Table 2**. The first two results are a full RCT (1:1 randomization ratio) and a hybrid trial (2:1 randomization ratio) with no borrowing for reference. As anticipated, these are unbiased estimates, however, a hybrid trial with no borrowing has less precision, demonstrated by an increased SE, due to the smaller sample size for the control arm.

An ideal method is expected to borrow information from historical trials to improve precision without introducing biases. When the weight parameter for the non-informative piece was set to 1, MAP utilized only the concurrent data, yielding results similar to a hybrid trial without borrowing. As the weight parameter decreased from 0.8 to 0.2, more information was borrowed from the historical trials, resulting in an increase in the effective sample size rate (ESSR) and a corresponding decrease in the standard error (SE). However, this came at the cost of higher biases and increased Type I errors. Similarly, mixed-effects models (MM) without covariates improved precision by sacrificing accuracy and Type I errors.

MAP and MM with no covariate adjustment do not take advantage of any existing confounder information directly. In contrast, other methods mitigated bias by including the confounders directly in either the PS model or the mixed-effects model. Given a correct model specification, MM showed the largest ESSR (238%), which indicated more than double the amount of information when compared to the concurrent data alone, and hence the biggest improvement in precision. Due to the most reduced SE, the mixed-effects model showed slight inflation in Type I error, though it controlled bias well. The performance of the mixed-effects model was followed by the two propensity score methods and the integrated method of PSS+PP. All three methods had ESSR greater than 100%, which implied strong borrowing from the historical trials. They all showed well-controlled Type I errors, though PSS+PP had a slightly increased bias. The PS-integrated methods of PSM/PSW + MAP also controlled Type I errors well. However, like MAP, the gain of precision depended on the choice of the weight parameter, where a smaller weight led to more borrowing from historical trials and a larger reduction in SE. Given the same weight parameter, PSW+MAP borrowed more information than PSM+MAP.

**Table 2** (right panel) also shows results for methods that include covariates when unmeasured confounders are present. It was assessed by removing the two synthetic covariates and leaving only baseline NT-proBNP. As expected, the performance of these methods deteriorated in this setting. The mixed-effects model, which outperformed other methods given the correct model, suffered the greatest impact. Borrowing was reduced as the ESSR went from 238% to 101%, and results showed higher bias and Type I error, indicating an incorrect borrowing behaviour. Since the baseline NT-proBNP only had a weak correlation with the outcome, the mixed model results with the misspecified model were similar to this approach with no covariates.

The model misspecification had trivial influences on the ESSR for PSM, PSW, and PSS+PP. However, large biases and Type I errors were observed for these three methods. The two PS-integrated methods, PSM/PSW+MAP, were also affected by the model misspecification. However, they can achieve a balance between the efficiency of borrowing and bias when the weight parameter is tuned to borrow a similar level of information as the full concurrent data (i.e., ESSR=50%). For example, PSM+MAP with a weight parameter of 0.5 lowered the SE from 0.082 to 0.069, similar to that with a fully randomized trial, and exhibited a slightly inflated Type I error of 0.062. Smaller weights on the historical data better controlled both bias and Type I error, although it also reduced the strength of borrowing.

*Simulation results*

**Table 3** shows the results of the simulation study with only one historical trial to borrow from in order to mimic the case study setting. It shows the Type I error on the left and power on the right, respectively. MAP could not enhance power given moderate to severe between-study heterogeneities. In the severe heterogeneity scenario, it appropriately disregards most of the historical data, introducing minimal bias (Supplementary Table 4 and Figure 3). In the moderate heterogeneity scenario, though MAP can slightly improve power by borrowing from the historical data, more bias and Type I error are present as more information is borrowed (**Table 3** and Supplementary Table 4).

Given a correct model specification, other methods that incorporate confounders directly enhanced power to at least the level of a full concurrent trial. While power tended to be higher with severe heterogeneities present, so were Type I errors. The mixed effects model and PSW performed the best with severe heterogeneities, controlling Type I errors and increasing power. However, under an incorrect model, the control of Type I error deteriorated for all these methods, especially under severe heterogeneities, with Type I errors increasing as more key confounders were excluded from the model. The mixed effects model and the two PS and MAP integrated methods demonstrated better Type I error control than other PS-integrated methods.

Figure 1 shows the results of the simulation study when multiple historical data sources were available for borrowing. In this setting, most methods showed well-controlled Type I errors under both correct and mildly incorrect model specifications (shown in the left panel of Supplementary Table 5). However, we observed both inflated Type I errors and increased biases for PSM, PSW, and PSS+PP when incorrect model specifications with more confounders omitted were present (Figure 1). The biases and Type I errors increased given more severe trial heterogeneities. Among these methods, only PSM, PSW, MM, and the integrated methods (i.e., PSS+PP, PSM+MAP, and PSW+MAP) can achieve or surpass the power levels observed in the full concurrent study (shown in the right panel of Supplementary Table 6). Note that we controlled the degree of borrowing by tuning the τ parameter rather than the weight parameter in MAP and MAP-integrated methods in this scenario. To achieve a similar ESSR to the full concurrent trial, an additional smaller τ parameter indicating a stronger borrowing was added to the severely heterogeneous scenario. Like what we observed previously in the single historical trial setting, the mixed-effects model outperformed other methods when at least partial confounding information was included in the model; otherwise, it can barely borrow any information like the MM without covariates (Figure 1 and Supplementary Table 6). Thanks to the flexible borrowing option via the τ parameter, PSW+MAP and PSM+MAP can adapt to optimal performance. In addition, they had better bias and Type I error control than PSM, PSW, and PSS+PP when achieving a similar enhancement in power.

**Discussion**

Hybrid-controlled trials with external data can potentially improve the efficiency of drug development. Our case study, along with an additional simulation study, demonstrates that statistical methods are crucial for addressing trial heterogeneity and effectively borrowing information to enhance the control arm. Methods like propensity score matching and weighting, the PS-integrated method of PSS+PP, and the mixed effect model outperformed others in terms of power enhancement in scenarios with moderate to severe heterogeneity between concurrent and historical trials. However, the success of these methods critically hinges on accurate model specifications, specifically that no unmeasured confounders are present. Failure to appropriately include all the important confounders may cause propensity score methods and PS-integrated methods to introduce unwanted bias that can lead to erroneous conclusions. When this assumption is hard to justify and concerns about unmeasured confounders arise, methods like PSM+MAP and PSW+MAP would be recommended because they have modelling mechanisms for both measured and unmeasured confounders, resulting in greater resistance to unmeasured confounding, while reaching a satisfactory balance in the bias-variance trade-off. Mixed modeling is also a good alternative if at least partial key confounders are assured to be included.

The performance of MAP and PS-integrated MAP depends on the choice of hyperparameters that govern how the external data is borrowed. Choices that result in weak borrowing can minimize bias but may compromise power enhancement, while choices that result in stronger borrowing might maximize power at the cost of increased bias, particularly in the presence of heterogeneity and model misspecification. As demonstrated in our simulations, a series of hyperparameters should ideally be evaluated through sensitivity analyses. A more objective solution involves selecting hyperparameters based on the ESSR, which measures the number of patients effectively added to the current trial. Liu et al. (2021) also propose to use ESS to tune the selection of hyperparameters that control the borrowing of an integrated approach of PSS+MAP. A desired number for ESSR can be roughly calculated from a hypothetical "full-size" concurrent study with 1:1 treatment allocation. For example, setting an ESSR target of approximately 50% could be a strategic choice, though this also necessitates considering factors like the availability of external data and between-study heterogeneity.

This research study has several limitations. Primarily, our simulation study only assesses the method's performance for continuous outcomes. However, binary and time-to-event outcomes, which are prevalent in oncology and rare disease settings, may require different analytical approaches. Some evaluated methods, like propensity score methods, are directly adaptable to binary or time-to-event data; some might need essential modifications. Additionally, although we delineated what should be matched in propensity score matching, we have not thoroughly discussed the estimands, i.e., what is being estimated. There is also a scarcity of literature addressing causal estimands within the framework of hybrid controlled designs (Lin et al. 2023).

In summary, the integration of external data holds considerable promise for enhancing underpowered control arms in hybrid-controlled trial designs. Our comprehensive analysis aims to deepen the understanding of statistical methodologies in this field and provide valuable insights for future enhancements in hybrid-controlled trial design. The optimal methods must account for the between-trial heterogeneities via both measured and unmeasured confounding. The PS and MAP integrated methods and mixed-effect models are generally more robust. However, given that no single method universally excels, researchers may have to consider multiple methods via sufficient simulations to assess their effectiveness under various scenarios. There is a pressing need for future research to develop a best practice guide that ensures efficient and robust hybrid-controlled trial designs.

## AUTHOR DISCLOSURE

Di Ran, Fanni Zhang, Binbing Yu, Kristine Broglio, and Sima Shahsavari are AstraZeneca employees. This research was conducted with the highest standards of academic integrity and objectivity. Each author has participated significantly in the conception, design, execution, or interpretation of the study and drafting of the paper. All authors have approved the final version of the manuscript for publication. The work presented in this paper was funded by AstraZeneca.

**Table 1** Listing of statistical methods with R packages that can be used for information borrowing in hybrid-controlled trials

| Methods | Abbrev. | Statistics | Handle Measured Confounding | Features | R packages |
|---|---|---|---|---|---|
| **Propensity score matching** | PSM | Frequentist | PS model | Mitigate known confounding without access to outcomes at the design stage; assume no unmeasured confounding. | MatchIt |
| **Inverse probability weighting using propensity score (with trimming)** | IPW | Frequentist | PS model | Mitigate known confounding without access to outcomes at the design stage; assume no unmeasured confounding. | stats |
| **Meta-analytic-predictive prior** | MAP | Bayesian | No | Mitigate trial heterogeneities due to confounding without assumption. Only need aggregate-level information rather than individual-level data. | RBesT |
| **Mixed-effect model** | MM | Frequentist | Covariate adjustment (optional) | Mitigate both known and unknown confounding via covariate modeling and random intercepts. | lme4 |
| **Propensity score (stratification)-integrated composite likelihood** | PSS+CL | PS integrated frequentist | PS model | Mitigate known confounding without access to outcomes at the design stage; address unknown confounding causing trial heterogeneities using CL | psrwe |
| **Propensity score (stratification)-integrated power prior** | PSS+PP | PS integrated Bayesian | PS model | Mitigate known confounding without access to outcomes at the design stage; address unknown confounding causing trial heterogeneities using PP | psrwe |
| **Propensity score mapping followed by MAP** | PSM+MAP | PS integrated Bayesian | PS model | Mitigate known confounding without access to outcomes at the design stage; address unknown confounding causing trial heterogeneities using MAP | MatchIt, RBesT |
| **MAP on aggregate information using inverse probability weighting using propensity score** | PSW+MAP | PS integrated Bayesian | PS model | Mitigate known confounding without access to outcomes at the design stage; address unknown confounding causing trial heterogeneities using MAP | stats, RBesT |

**Table 2** Summary of Method Assessment in Case Study based on DAPA-HF

| | Correct model with all confounders | | | | | Incorrect model with baseline NT-proBNP only | | | |
|---|---|---|---|---|---|---|---|---|---|
| **Methods** | **Bias** | **SE** | **Type I error** | **ESSR (%)** | **Methods** | **Bias** | **SE** | **Type I error** | **ESSR (%)** |
| unadj.rc | -0.006 | 0.082 | 0.040 | 0 | unadj.rc | - | - | - | - |
| unadj.fc | -0.004 | 0.067 | 0.051 | 50 | unadj.fc | - | - | - | - |
| MAP(.2) | -0.070 | 0.065 | 0.243 | 60 | MAP(.2) | - | - | - | - |
| MAP(.5) | -0.054 | 0.072 | 0.152 | 29 | MAP(.5) | - | - | - | - |
| MAP(.8) | -0.034 | 0.078 | 0.082 | 10 | MAP(.8) | - | - | - | - |
| MAP(1) | -0.007 | 0.082 | 0.044 | 1 | MAP(1) | - | - | - | - |
| MM | 0.000 | 0.045 | 0.060 | 238 | MM | -0.028 | 0.058 | 0.136 | 101 |
| MM.nc | -0.036 | 0.063 | 0.152 | 72 | MM.nc | - | - | - | - |
| PSM | -0.003 | 0.056 | 0.046 | 115 | PSM | -0.044 | 0.057 | 0.121 | 106 |
| PSW | 0.007 | 0.055 | 0.033 | 125 | PSW | -0.049 | 0.053 | 0.143 | 137 |
| PSS+PP | -0.012 | 0.052 | 0.038 | 152 | PSS+PP | -0.061 | 0.053 | 0.189 | 143 |
| PSM+MAP(.2) | -0.008 | 0.061 | 0.035 | 81 | PSM+MAP(.2) | -0.039 | 0.063 | 0.096 | 68 |
| PSM+MAP(.5) | -0.008 | 0.065 | 0.034 | 58 | PSM+MAP(.5) | -0.032 | 0.069 | 0.062 | 43 |
| PSM+MAP(.8) | -0.009 | 0.072 | 0.029 | 31 | PSM+MAP(.8) | -0.021 | 0.075 | 0.042 | 21 |
| PSM+MAP(1) | -0.009 | 0.082 | 0.045 | 2 | PSM+MAP(1) | -0.007 | 0.082 | 0.047 | 1 |
| PSW+MAP(.2) | 0.007 | 0.056 | 0.028 | 116 | PSW+MAP(.2) | -0.051 | 0.059 | 0.136 | 92 |
| PSW+MAP(.5) | 0.004 | 0.061 | 0.022 | 83 | PSW+MAP(.5) | -0.042 | 0.066 | 0.086 | 55 |
| PSW+MAP(.8) | 0.000 | 0.069 | 0.019 | 43 | PSW+MAP(.8) | -0.028 | 0.074 | 0.052 | 25 |
| PSW+MAP(1) | -0.007 | 0.082 | 0.045 | 1 | PSW+MAP(1) | -0.007 | 0.082 | 0.044 | 1 |

Unadj, unadjusted; rc, reduced concurrent trial (with half concurrent controls); fc, full concurrent trial (with full concurrent controls); ESSR, effective sample size rate; PSM, propensity score matching; PSW, propensity score weighting; PSS, propensity score stratification; PP, power prior; MAP, meta-analytic-predictive prior; MM, mixed effect model; MM.nc, mixed effect model with no covariates.

**Table 3** Summary of Simulation with Single Historical Trials

| | Type 1 error | | | | | | Power | | | | | |
|---|---|---|---|---|---|---|---|---|---|---|---|---|
| | Moderate | | | Severe | | | Moderate | | | Severe | | |
| | Model1 | Model2 | Model3 | Model1 | Model2 | Model3 | Model1 | Model2 | Model3 | Model1 | Model2 | Model3 |
| **unadj.rc** | 0.051 | | | 0.061 | | | 75.7 | | | 78.5 | | |
| **unadj.fc** | 0.052 | | | 0.062 | | | 89.6 | | | 92.3 | | |
| **MAP(.2)** | 0.305 | | | 0.063 | | | 82.5 | | | 81.6 | | |
| **MAP(.5)** | 0.179 | | | 0.065 | | | 79.5 | | | 81.3 | | |
| **MAP(.8)** | 0.101 | | | 0.063 | | | 77.5 | | | 81.3 | | |
| **MAP(1)** | 0.050 | | | 0.067 | | | 78.0 | | | 81.2 | | |
| **MM.nc** | 0.092 | | | 0.065 | | | 82.9 | | | 80.9 | | |
| **MM** | 0.058 | 0.076 | 0.122 | 0.044 | 0.095 | 0.053 | 89.3 | 87.5 | 84.4 | 97.0 | 95.7 | 90.3 |
| **PSM** | 0.083 | 0.124 | 0.308 | 0.121 | 0.399 | 0.962 | 95.6 | 98.5 | 99.8 | 98.2 | 100.0 | 100.0 |
| **PSW** | 0.037 | 0.061 | 0.283 | 0.049 | 0.272 | 0.950 | 94.4 | 98.6 | 99.9 | 97.1 | 99.9 | 100.0 |
| **PSS+PP** | 0.054 | 0.105 | 0.324 | 0.118 | 0.551 | 0.995 | 99.3 | 99.9 | 100.0 | 100.0 | 100.0 | 100.0 |
| **PSM+MAP(.2)** | 0.060 | 0.095 | 0.206 | 0.088 | 0.264 | 0.141 | 94.6 | 96.8 | 94.5 | 97.0 | 93.5 | 83.4 |
| **PSM+MAP(.5)** | 0.048 | 0.073 | 0.141 | 0.077 | 0.200 | 0.102 | 92.8 | 94.6 | 91.4 | 94.7 | 90.9 | 83.7 |
| **PSM+MAP(.8)** | 0.041 | 0.059 | 0.094 | 0.060 | 0.121 | 0.079 | 89.9 | 91.1 | 86.3 | 92.5 | 87.9 | 83.4 |
| **PSM+MAP(1)** | 0.056 | 0.055 | 0.050 | 0.069 | 0.067 | 0.067 | 80.3 | 80.1 | 79.2 | 86.0 | 85.1 | 83.9 |
| **PSW+MAP(.2)** | 0.046 | 0.072 | 0.222 | 0.087 | 0.258 | 0.128 | 95.8 | 97.4 | 94.2 | 96.9 | 92.1 | 80.5 |
| **PSW+MAP(.5)** | 0.041 | 0.053 | 0.154 | 0.071 | 0.182 | 0.089 | 93.5 | 95.2 | 91.7 | 94.7 | 88.5 | 81.2 |
| **PSW+MAP(.8)** | 0.026 | 0.038 | 0.104 | 0.063 | 0.122 | 0.077 | 88.9 | 91.4 | 85.9 | 91.1 | 84.2 | 80.9 |
| **PSW+MAP(1)** | 0.049 | 0.049 | 0.051 | 0.066 | 0.064 | 0.064 | 78.2 | 78.0 | 77.9 | 82.3 | 81.5 | 81.9 |

The colours represent the magnitude of Type I error (ranging from light yellow to red in the left panel) and power (ranging from white to green in the right panel) from low to high. The results are shown for two scenarios, simulating moderate and severe among-study heterogeneities, respectively. Model1: the correct model specifications; Model2: the incorrect model specification, achieved by removing one confounder when confounders are required; Model3 the incorrect model specification, achieved by removing three confounders when confounders are required. Unadj, unadjusted; rc, reduced concurrent trial (with half concurrent controls); fc, full concurrent trial (with full concurrent controls); ESSR, effective sample size rate; PSM, propensity score matching; PSW, propensity score weighting; PSS, propensity score stratification; PP, power prior; MAP, meta-analytic-predictive prior; MM, mixed effect model; MM.nc, mixed effect model with no covariates.

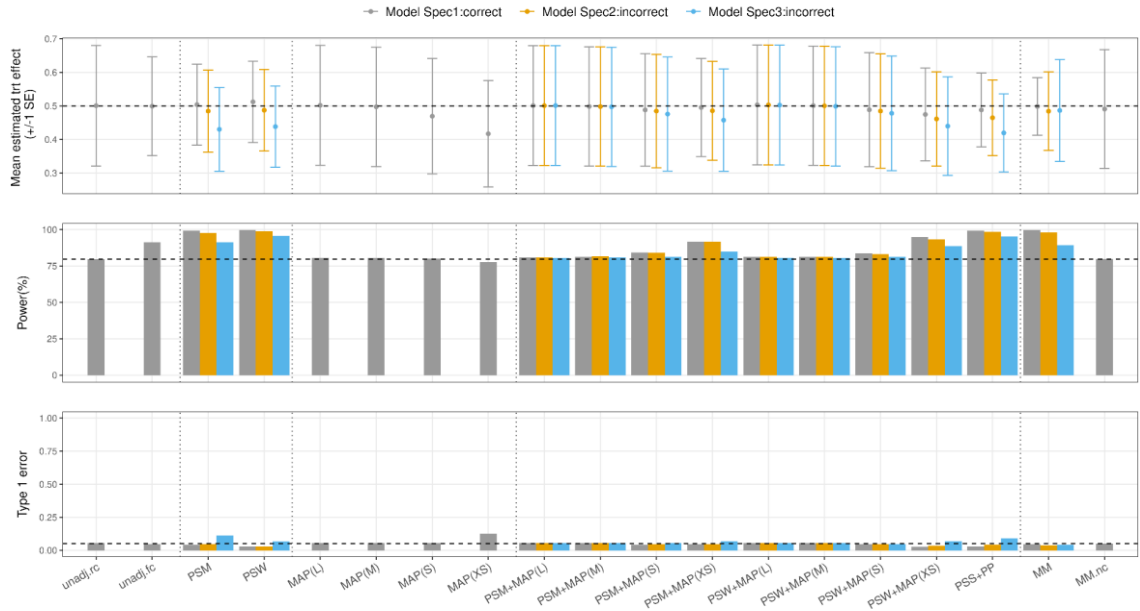

**Fig 1** Simulation results demonstrate scenarios involving multiple historical trials and severe heterogeneities between the concurrent and historical trials. The upper, middle, and bottom panels display the mean estimated treatment effect along with the corresponding ± 1 estimated standard error, power under the alternative hypothesis, and Type I error under the null hypothesis, respectively. Dashed horizontal lines represent the true effect size ($\theta_{trt} = 0.5$), power for the unadjusted model with concurrent treated and control data (approximately 75%), and Type I error at a level of 0.5. The results of different model specifications are shown in different colours: (grey) unadjusted model without any confounders or correct model specifications with all six confounders, (orange) incorrect model specification 1, which includes five confounders and omits the confounder $x_4$, and (blue) incorrect model specification 2, which includes three confounders ($x_1, x_2, x_3$). Unadj, unadjusted; rc, reduced concurrent trial (with half concurrent controls); fc, full concurrent trial (with full concurrent controls); PSM, propensity score matching; PSW, propensity score weighting; PSS, propensity score stratification; PP, power prior; MAP, meta-analytic-predictive prior; L, M, S, XS, large, medium, small and extra small weight for MAP; MM, mixed effect model; MM.nc, mixed effect model with no covariates.

Supplementary Materials

Supplementary materials for

# A Comparative Evaluation of Statistical Methods in Hybrid Controlled Trials

Di Ran[a]*, Fanni Zhang[a], Sima Shahsavari[b], Kristine Broglio[a], Alasdair Henderson[c], and Binbing Yu[a]

[a]*Oncology Biometrics Statistical Innovation, AstraZeneca, Gaithersburg, USA;* [b]*Late CVRM Biometrics, AstraZeneca, City, Gothenburg, Sweden;* [c]*School of Cardiovascular and Metabolic Health, University of Glasgow, UK*

*Corresponding author: Di Ran, di.ran@astrazeneca.com, Oncology Biometrics Statistical Innovation, AstraZeneca, 1 MedImmune Way, Gaithersburg, Maryland, 20878, USA

**PS weighting/matching integrated with MAP**

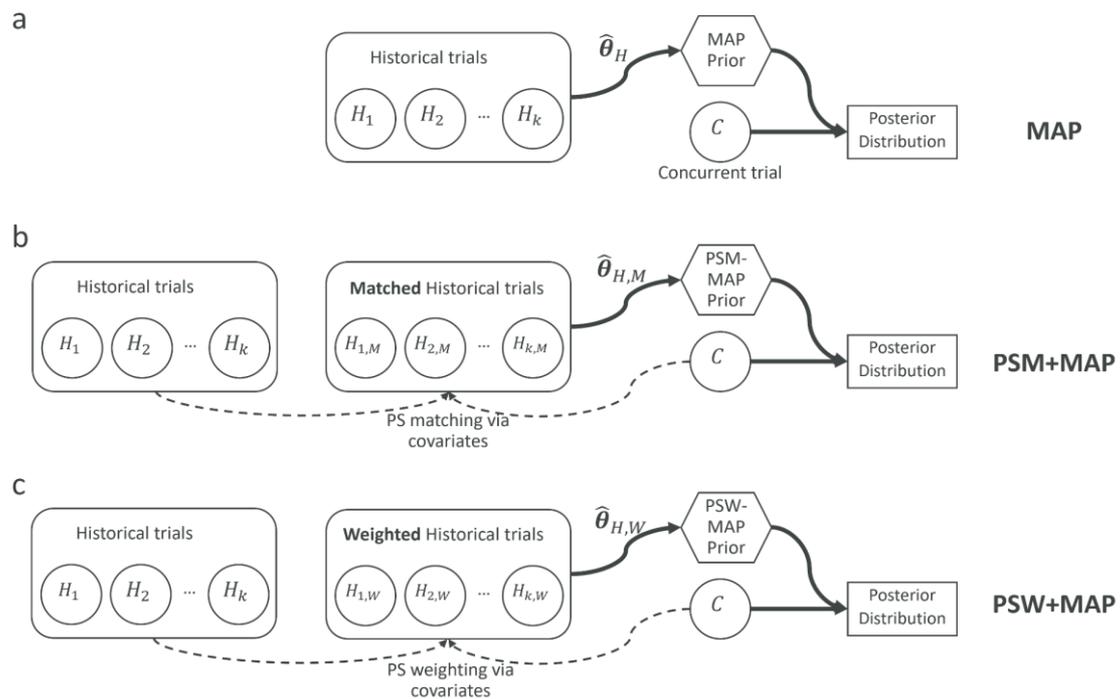

**Fig 2** Illustration of the Meta-analytic predictive prior (MAP) and propensity-score integrated MAP approaches, including propensity-score matching followed by MAP (PSM+MAP) and PS inverse probability weighting with MAP (PSW+MAP). The dashed lines on panels B and C illustrate that the propensity scores of being in the concurrent trial are calculated for the subjects in the concurrent and historical trials. Based on these propensity scores, matched subjects are selected in PSM+MAP, and inverse probability weighted subjects are used in PSW+MAP, respectively, to compute the MAP prior.

**Details of the Case Study**

*Create synthetic covariates in case study*

To demonstrate the application of the methods, we utilize a case study based on patient-level data from the DAPA-HF clinical trial which assessed the safety and efficacy of dapagliflozin in reducing the risk of worsening heart failure or cardiovascular death in patients with heart failure, irrespective of their diabetes status (McMurray, Solomon et al. 2019). Our analytical focus for this case study is on the change from baseline to 8 months in the log-transformed N-terminal pro-B-type natriuretic peptide (NT-proBNP) levels. NT-proBNP is secreted by cardiac myocytes in response to increased cardiac wall stress, a key indicator of heart failure, making it a critical diagnostic and prognostic biomarker. We limit our analysis to patient-level data from the placebo arm, with data available for 1,892 patients after excluding those with missing outcomes, incomplete baseline characteristics (encompassing eighteen variables), and those who withdrew consent due to personal reasons or data governance policies in specific countries.

Through literature searches and consultations with experts, we identify 18 confounding variables including age, sex, body mass index (BMI), ethnicity, race, country of origin, NYHA functional classification, estimated glomerular filtration rate (GFR), systolic and diastolic blood pressures, pulse rate, left ventricular ejection fraction, baseline NT-proBNP levels, type 2 diabetes status, history of atrial fibrillation, HbA1c, and ECG mean QRS duration. Among these factors, however, only the baseline NT-proBNP exhibits a weak correlation of $\rho = -0.22$ with the outcome, which stands out not only for its statistical significance but also for its practical relevance. In contrast, other variables have negligible correlation coefficients despite significant p-values. To ensure the confounding mitigation model works, we introduce two synthetic covariates, $x_1$ and

$x_2$, which demonstrate stronger correlations to the outcome with $\rho(x_1, y) = 0.5$ and $\rho(x_2, y) = 0.4$. The generation of these covariates is achieved using the `create.cor.var` function, detailed in the source R script `rcode_sim_single_source.R`. The baseline NT-proBNP and these two synthetic covariates are incorporated into propensity score calculations using a logistic model with pre-defined coefficients.

For our "concurrent" trial emulation, we first shuffle the total of 1,892 patients in the DAPA-HF control group and select 600 subjects using a Bernoulli distribution where the probability of success (i.e., being in the "concurrent" trials) is determined by the propensity score. In detail, if a patient has a propensity score that exceeds the average, we assign this patient a higher probability of 0.65 of being in the "concurrent" trial; otherwise, we give a lower probability of 0.35. This patient will be assigned to the "concurrent" trial if a successful outcome is observed from the Bernoulli sampling. We repeat this process until we reach the target number (n=600) of patients. These "concurrent" patients are then randomly divided into treatment and control subgroups equally, while the remainder form a historical control group.

We assess the method performance through the single historical control borrowing framework as detailed in the simulation section. Two model specifications are examined: a correct model comprising the baseline NT-proBNP, $x_1$ and $x_2$, and an incorrect model relying on the baseline NT-proBNP alone. We iterated the covariate creation and patient sampling 1000 times to ensure consistency and reduce the variability of our results.

**Details of the simulation study**

*Simulation of data with single or multiple historical controls*

Simulations were conducted to examine how the selected methods perform in the hybrid-controlled trial setting under two scenarios: one single historical trial and multiple historical trials. Since existing work has explored the scenarios where the external data is homogenous with the concurrent study or mild heterogeneity is present, our simulation focused exclusively on scenarios exhibiting moderate to severe heterogeneity between the concurrent study and external data.

For the single historical control scenario, we simulated data for 1,200 subjects, each with six independent variables. These 6 variables ($x_1$ to $x_6$) were drawn from normal distributions. The treatment status was generated independently from the covariates, with a 1:1 ratio between treatment and control arms, i.e., $z_i \sim \text{Bernoulli}(0.5)$. The log odds of being in the concurrent trial were modeled as a linear combination of the six variables, where we fixed the values of the coefficients to create an association between the covariates and the probability of being in the concurrent trial.

$$\text{Logit}(p_{i,\text{trial}}) = \beta_0 + \sum_{l=1}^{6} \beta_l x_{il}$$

The intercept parameter ($\beta_0$) was tuned to assign roughly 400 subjects in the concurrent trial and 800 subjects in a relatively larger historical trial. The calculated log odds were then translated into subject-specific probabilities and used to generate each patient's trial assignment from a Bernoulli distribution.

$$T_i \sim \text{Bernoulli}(p_{i,\text{trial}})$$

The outcome for each of the 1200 subjects was then randomly generated conditional on the treatment assignment status and the six variables where again the coefficient values were fixed in order to create an association between the covariates and the outcome,

$$y_i = \alpha_0 + \theta_{treat} z_i + \sum_{l=1}^{6} \alpha_l x_{il} + e_i$$

where $e_i$ represents the random errors, assumed to be independently and identically distributed (i.i.d.) with $e_i \sim N(0, 1)$. To assess the performance of the selected methods under varying levels of heterogeneity between the outcomes of concurrent and historical controls, we implemented two sets of coefficients, one set to create moderate heterogeneity and one set to create severe heterogeneity across the trials. For a severe heterogeneity setting, the coefficients were set as follows: $\beta_l = \alpha_l = 0.5$ for $l = 1, \ldots, 6$, $\beta_0 = -0.9$, $\alpha_0 = 1$, and $\theta_{treat} = 0.5$ (or 0 for assessing type-I error). For a moderate heterogeneity setting, we used $\beta_l = 0.3, \alpha_l = 0.2$ for $l = 1, \ldots, 6$, $\beta_0 = -0.78$, $\alpha_0 = 1$, and $\theta_{treat} = 0.35$ (or 0 for assessing type-I error). Both configurations are designed to provide a rough 90% power for the crude model based on the full concurrent dataset. For the scenario involving multiple historical controls, we used a multinomial logit model to simulate trial assignments. The probability of being in the concurrent trial or one of three historical controls ($j = 1, 2, 3$) was defined as:

$$p_{i,concurrent} = \frac{1}{1+\sum_{j=1}^{3} \exp(\beta_{j0}+\sum_{l=1}^{6} \beta_{jl}x_{il})},$$

$$p_{ij} = p_{concurrent} \times \exp(\beta_{j0} + \sum_{l=1}^{6} \beta_{jl}x_{il}).$$

As with the single control scenario, we established one moderate and one severe setting for the "multiple-historical controls" to reflect varying degrees of heterogeneity. We chose coefficient sets aiming for approximately 90% power for the crude model on the full concurrent dataset. The parameters are summarized in the following table.

| Coefficients | Single historical control | | Multiple historical controls, k=3 | |
| --- | --- | --- | --- | --- |
| | Moderate | Severe | Moderate | Severe |
| $\alpha_0$ | 1 | 1 | 1.2 | 1 |
| $\alpha_l$ | 0.2 | 0.5 | 0.5 | 0.5 |
| $\beta_0$ | -0.78 | -0.9 | $\beta_{10}=0.8, \beta_{20}=-1, \beta_{30}=-0.7$ | $\beta_{10}=-1, \beta_{20}=-0.1, \beta_{30}=0.2$ |
| $\beta_l$ | 0.3 | 0.5 | $\beta_{1l}=0.1, \beta_{2l}=0, \beta_{3l}=-0.1$ | $\beta_{1l}=0.1, \beta_{2l}=0.4, \beta_{3l}=-0.2$ |
| $\theta_{treat}$ | 0.35 | 0.5 | 0.5 | 0.5 |

A total of 1,600 subjects were simulated with approximately 25% (i.e., 400) allocated to the concurrent trial and the remainder distributed among the three historical controls.

In each simulation, we draw from the virtual subjects a fully randomized trial with a 1:1 ratio between treatment and control arms. This is used to estimate the standard error of the treatment effect that would ideally be achieved for a complete RCT. For the hybrid control setting, we then randomly drop half of the subjects on the control arm, creating a 2:1 ratio between treated and control arms. We also calculate the standard error of the treatment effect with only these concurrent trial patients to capture the impact of decreased sample size. Treatment effects are estimated for the full and reduced trials, unadjusted for any covariates, and are unbiased as they are purely randomized comparisons. This provides two benchmarks against which to measure the performance of borrowing in historical trials. Then, finally, we use the virtual subjects in the historical trials to augment the concurrent controls, and estimates are captured for each analysis method. The data generation process was replicated 2,000 times for each scenario and setting to ensure reliability and statistical validity in our findings.

*Model specifications for confounding mitigation in simulation*

To evaluate the robustness of the methods to unmeasured confounders, we tested them under various conditions in which the analysis models may be misspecified due to the absence of key confounders. To this end, we set up three distinct model specifications:

- Correct Model: Includes all six confounders $(x_1, \ldots, x_6)$, representing an ideal scenario where all relevant variables are accounted for.
- Incorrect Model 1: Omits confounder $x_4$ from the perfect model, thereby including only five confounders $(x_1, x_2, x_3, x_5, x_6)$, to simulate a minor misspecification.
- Incorrect Model 2: Further reduces the number of confounders by omitting two additional variables ($x_5$ and $x_6$) from Incorrect Model 1, resulting in a model with just three confounders $(x_1, x_2, x_3)$, to simulate more severe misspecification.

*Method specifications in simulation*

For propensity score-based methods, logistic regressions were used to compute the propensity score ($e_i$), representing the probability of each subject being in the concurrent trial. PS matching identified matched historical controls to the entire concurrent trial (both treated and control groups) with replacement using the R package *MatchIt*. A caliper width of 0.2 was used to limit the subjects to be paired. Cluster robust standard error was computed by using the R package *miceadds*. PS weighting used ATT weights for which all the subjects of the concurrent trial have weights equal to 1 and subjects of the historical control have weights of $\frac{e_i}{1-e_i}$. We applied symmetric trimming to remove extremely small or large weights, specifically those below 0.05 and above 20. The robust standard error was computed by using the R package *survey*. The mixed-effect model allowed random intercepts between concurrent and historical controls. We used the R package *lme4* for the two mixed-effect models with different model specifications: one with treatment as the sole covariate and the other incorporating both treatment and additional variables from the analysis models. For MAP, the R package *RBesT* was used for the approximation and robustification of the prior as well as the posterior estimation. We controlled the amount of information borrowed using a series of weight parameters (0.2, 0.5, 0.8, 1), indicating most to no borrowing. In the multiple historical control scenario, both the weight parameter and the tau prior can be used to regulate the extent of information borrowed. We have set the weight parameter at a constant value of 0.5, allowing us to focus on adjusting the tau prior to controlling the level of information borrowing. The tau prior is varied from larger to smaller values, which corresponds to borrowing less to more information, respectively. The observed variability between trials serves as the basis for selecting this parameter. The larger and smaller choices are 10 times and 1/10 of that empirical choice. R package *psrwe* is used for both PSS+PP and PSS+CL with five strata and all default values for other parameters. Note that during the implementation of the PSS+CL method, we observed unreasonably high values for the standard deviation (`sd_theta`). Consequently, we revised its calculation in the function `get_cl_stratum` to reflect that the standard deviation of the sampling distribution of the mean is more accurately termed the standard error of the mean. The updated function, `get_cl_stratum_new`, is detailed in the `rcode_update_psrwe_source.R` script.

# Additional simulation results

**Table 4** Summary of method assessment for simulated data with a single historical trial

| Methods | Moderate heterogeneities | | | | | Severe heterogeneities | | | | |
|---|---|---|---|---|---|---|---|---|---|---|
| | Bias | RB(%) | Type1error | Power(%) | ESSR(%) | Bias | RB(%) | Type1error | Power(%) | ESSR(%) |
| unadj.rc | 0.012 | 3.3 | 0.051 | 75.7 | 0.0 | -0.002 | -0.4 | 0.061 | 78.5 | 0.0 |
| unadj.fc | 0.006 | 1.7 | 0.052 | 89.6 | 49.7 | -0.003 | -0.5 | 0.062 | 92.3 | 49.4 |
| MAP(.2) | 0.138 | 39.5 | 0.305 | 82.5 | -7.2 | 0.016 | 3.1 | 0.063 | 81.6 | 0.6 |
| MAP(.5) | 0.094 | 26.8 | 0.179 | 79.5 | -11.9 | 0.015 | 3.1 | 0.065 | 81.3 | 0.7 |
| MAP(.8) | 0.057 | 16.2 | 0.101 | 77.5 | -9.2 | 0.016 | 3.1 | 0.063 | 81.3 | 0.7 |
| MAP(1) | 0.023 | 6.7 | 0.050 | 78.0 | 1.0 | 0.016 | 3.1 | 0.067 | 81.2 | 0.6 |
| MM.nc | 0.054 | 15.3 | 0.092 | 82.9 | 15.5 | 0.017 | 3.4 | 0.065 | 80.9 | -0.6 |
| *Model specification1: correct model with all confounders* | | | | | | | | | | |
| PSM | 0.000 | 0.1 | 0.083 | 95.6 | 116.9 | 0.004 | 0.9 | 0.121 | 98.2 | 173.2 |
| PSW | 0.000 | 0.1 | 0.037 | 94.4 | 84.4 | 0.046 | 9.2 | 0.049 | 97.1 | 59.4 |
| PSM+MAP(.2) | 0.014 | 4.1 | 0.060 | 94.6 | 77.0 | 0.034 | 6.8 | 0.088 | 97.0 | 83.5 |
| PSM+MAP(.5) | 0.018 | 5.0 | 0.048 | 92.8 | 56.3 | 0.035 | 7.0 | 0.077 | 94.7 | 62.8 |
| PSM+MAP(.8) | 0.023 | 6.5 | 0.041 | 89.9 | 31.3 | 0.038 | 7.6 | 0.060 | 92.5 | 36.8 |
| PSM+MAP(1) | 0.033 | 9.5 | 0.056 | 80.3 | 1.0 | 0.046 | 9.2 | 0.069 | 86.0 | 2.0 |
| PSW+MAP(.2) | 0.003 | 0.8 | 0.046 | 95.8 | 92.8 | 0.047 | 9.4 | 0.087 | 96.9 | 90.1 |
| PSW+MAP(.5) | 0.006 | 1.6 | 0.041 | 93.5 | 69.5 | 0.042 | 8.4 | 0.071 | 94.7 | 65.9 |
| PSW+MAP(.8) | 0.011 | 3.2 | 0.026 | 88.9 | 39.7 | 0.034 | 6.8 | 0.063 | 91.1 | 36.9 |
| PSW+MAP(1) | 0.024 | 6.8 | 0.049 | 78.2 | 0.9 | 0.019 | 3.8 | 0.066 | 82.3 | 0.8 |
| PSS+CL | 0.026 | 7.6 | 0.000 | 11.1 | -70.6 | 0.076 | 15.2 | 0.000 | 68.7 | -53.6 |
| PSS+PP | 0.027 | 7.6 | 0.054 | 99.3 | 112.3 | 0.084 | 16.9 | 0.118 | 100.0 | 198.6 |
| MM | 0.007 | 2.0 | 0.058 | 89.3 | 114.6 | -0.001 | -0.3 | 0.044 | 97.0 | 259.9 |
| *Model specification2: incorrect model omitting one confounder* | | | | | | | | | | |
| PSM | 0.044 | 12 | 0.124 | 98.5 | 113.4 | 0.182 | 36.4 | 0.399 | 100.0 | 153.9 |
| PSW | 0.046 | 13 | 0.061 | 98.6 | 90.9 | 0.194 | 38.9 | 0.272 | 99.9 | 66.2 |
| PSM+MAP(.2) | 0.049 | 14 | 0.095 | 96.8 | 75.5 | 0.161 | 32.1 | 0.264 | 93.5 | 42.2 |
| PSM+MAP(.5) | 0.045 | 13 | 0.073 | 94.6 | 54.9 | 0.131 | 26.1 | 0.200 | 90.9 | 24.6 |
| PSM+MAP(.8) | 0.040 | 12 | 0.059 | 91.1 | 30.3 | 0.095 | 18.9 | 0.121 | 87.9 | 11.5 |
| PSM+MAP(1) | 0.033 | 9 | 0.055 | 80.1 | 0.8 | 0.042 | 8.5 | 0.067 | 85.1 | 1.7 |
| PSW+MAP(.2) | 0.045 | 13 | 0.072 | 97.4 | 91.4 | 0.155 | 31.0 | 0.258 | 92.1 | 38.2 |
| PSW+MAP(.5) | 0.041 | 12 | 0.053 | 95.2 | 67.9 | 0.119 | 23.7 | 0.182 | 88.5 | 18.9 |
| PSW+MAP(.8) | 0.035 | 10 | 0.038 | 91.4 | 38.6 | 0.076 | 15.3 | 0.122 | 84.2 | 6.7 |
| PSW+MAP(1) | 0.024 | 7 | 0.049 | 78.0 | 0.8 | 0.017 | 3.5 | 0.064 | 81.5 | 0.8 |
| PSS+CL | 0.066 | 19 | 0.000 | 21.7 | -70.8 | 0.221 | 44.2 | 0.004 | 95.4 | -56.3 |
| PSS+PP | 0.068 | 19 | 0.105 | 99.9 | 108.9 | 0.234 | 46.8 | 0.551 | 100.0 | 162.6 |
| MM | 0.031 | 9 | 0.076 | 87.5 | 104.0 | 0.046 | 9.2 | 0.095 | 95.7 | 122.4 |
| *Model specification3: incorrect model omitting three confounders* | | | | | | | | | | |
| PSM | 0.129 | 37 | 0.308 | 99.8 | 106.7 | 0.489 | 97.8 | 0.962 | 100.0 | 124.4 |
| PSW | 0.130 | 37 | 0.283 | 99.9 | 101.4 | 0.490 | 97.9 | 0.950 | 100.0 | 85.3 |
| PSM+MAP(.2) | 0.106 | 30 | 0.206 | 94.5 | 48.7 | 0.099 | 19.8 | 0.141 | 83.4 | -16.6 |
| PSM+MAP(.5) | 0.087 | 25 | 0.141 | 91.4 | 29.9 | 0.066 | 13.1 | 0.102 | 83.7 | -10.3 |
| PSM+MAP(.8) | 0.063 | 18 | 0.094 | 86.3 | 14.1 | 0.047 | 9.3 | 0.079 | 83.4 | -4.7 |
| PSM+MAP(1) | 0.028 | 8 | 0.050 | 79.2 | 0.6 | 0.033 | 6.7 | 0.067 | 83.9 | 0.9 |
| PSW+MAP(.2) | 0.113 | 32 | 0.222 | 94.2 | 57.4 | 0.075 | 15.0 | 0.128 | 80.5 | -18.5 |
| PSW+MAP(.5) | 0.093 | 27 | 0.154 | 91.7 | 33.6 | 0.044 | 8.7 | 0.089 | 81.2 | -11.4 |
| PSW+MAP(.8) | 0.066 | 19 | 0.104 | 85.9 | 15.1 | 0.026 | 5.3 | 0.077 | 80.9 | -4.9 |
| PSW+MAP(1) | 0.024 | 7 | 0.051 | 77.9 | 0.8 | 0.016 | 3.3 | 0.064 | 81.9 | 0.7 |
| PSS+CL | 0.142 | 41 | 0.000 | 48.0 | -71.2 | 0.505 | 101.1 | 0.342 | 100.0 | -60.0 |
| PSS+PP | 0.144 | 41 | 0.324 | 100.0 | 103.4 | 0.534 | 106.9 | 0.995 | 100.0 | 118.2 |
| MM | 0.058 | 16 | 0.122 | 84.4 | 60.3 | 0.024 | 4.7 | 0.053 | 90.3 | 32.7 |

Unadj, unadjusted; rc, reduced concurrent trial (with half concurrent controls); fc, full concurrent trial (with full concurrent controls); ESSR, effective sample size rate; PSM, propensity score matching; PSW, propensity score weighting; PSS, propensity score stratification; PP, power prior; CL, composite likelihood; MAP, meta-analytic-predictive prior; MM, mixed effect model; MM.nc, mixed effect model with no covariates.

**Table 5** Summary of method assessment for simulated data with multiple historical trials

| Methods | Moderate heterogeneities | | | | | Severe heterogeneities | | | | |
|---|---|---|---|---|---|---|---|---|---|---|
| | Bias | RB(%) | Type1error | Power(%) | ESSR(%) | Bias | RB(%) | Type1error | Power(%) | ESSR(%) |
| unadj.rc | 0.003 | 0.6 | 0.040 | 72.8 | 0.0 | 0.001 | 0.1 | 0.054 | 79.7 | 0.0 |
| unadj.fc | 0.001 | 0.3 | 0.053 | 88.5 | 49.9 | -0.001 | -0.2 | 0.048 | 91.4 | 49.5 |
| MAP(L) | -0.001 | -0.1 | 0.033 | 75.5 | 7.3 | 0.002 | 0.4 | 0.056 | 80.7 | 1.2 |
| MAP(M) | -0.007 | -1.5 | 0.034 | 78.2 | 13.6 | -0.003 | -0.5 | 0.054 | 80.6 | 2.1 |
| MAP(S) | -0.064 | -12.8 | 0.078 | 77.6 | 40.6 | -0.031 | -6.1 | 0.053 | 80.1 | 9.2 |
| MAP(XS) | | | | | | -0.083 | -16.6 | 0.125 | 77.8 | 29.1 |
| MM.nc | -0.008 | -1.5 | 0.027 | 78.8 | 20.5 | -0.009 | -1.9 | 0.052 | 79.9 | 2.9 |
| | Model specification1: correct model with all confounders | | | | | | | | | |
| PSM | 0.003 | 0.5 | 0.045 | 97.7 | 129.5 | 0.004 | 0.8 | 0.042 | 99.3 | 121.4 |
| PSW | 0.001 | 0.3 | 0.015 | 98.1 | 128.5 | 0.012 | 2.4 | 0.028 | 99.6 | 119.6 |
| PSM+MAP(L) | 0.009 | 1.8 | 0.035 | 77.3 | 9.0 | 0.001 | 0.2 | 0.054 | 80.8 | 1.4 |
| PSM+MAP(M) | 0.013 | 2.6 | 0.025 | 81.5 | 16.3 | -0.002 | -0.3 | 0.053 | 81.2 | 2.7 |
| PSM+MAP(S) | 0.009 | 1.8 | 0.026 | 90.0 | 50.6 | -0.012 | -2.4 | 0.042 | 84.3 | 14.5 |
| PSM+MAP(XS) | | | | | | -0.005 | -0.9 | 0.046 | 91.8 | 51.3 |
| PSW+MAP(L) | 0.016 | 3.2 | 0.034 | 78.5 | 8.5 | 0.003 | 0.6 | 0.056 | 81.2 | 1.1 |
| PSW+MAP(M) | 0.022 | 4.5 | 0.028 | 83.1 | 16.3 | 0.001 | 0.2 | 0.054 | 81.4 | 2.2 |
| PSW+MAP(S) | 0.013 | 2.7 | 0.009 | 93.9 | 67.2 | -0.011 | -2.3 | 0.045 | 83.8 | 11.9 |
| PSW+MAP(XS) | | | | | | -0.025 | -5.1 | 0.027 | 95.0 | 68.5 |
| PSS+CL | -0.015 | -3.0 | 0.000 | 28.9 | -50.3 | -0.012 | -2.3 | 0.000 | 35.0 | -55.9 |
| PSS+PP | -0.016 | -3.1 | 0.029 | 99.0 | 180.1 | -0.012 | -2.4 | 0.030 | 99.3 | 164.8 |
| MM | -0.004 | -0.8 | 0.043 | 98.8 | 396.3 | -0.002 | -0.3 | 0.046 | 99.6 | 335.2 |
| | Model specification2: incorrect model omitting one confounder | | | | | | | | | |
| PSM | -0.020 | -4.0 | 0.057 | 95.6 | 124.3 | -0.015 | -3.1 | 0.048 | 97.7 | 116.4 |
| PSW | -0.020 | -4.1 | 0.025 | 96.8 | 128.9 | -0.013 | -2.5 | 0.028 | 98.8 | 120.2 |
| PSM+MAP(L) | 0.006 | 1.1 | 0.031 | 77.8 | 8.7 | 0.001 | 0.2 | 0.053 | 81.0 | 1.3 |
| PSM+MAP(M) | 0.007 | 1.4 | 0.027 | 80.8 | 15.6 | -0.002 | -0.4 | 0.055 | 81.8 | 2.5 |
| PSM+MAP(S) | -0.004 | -0.7 | 0.031 | 88.9 | 48.1 | -0.015 | -3.1 | 0.047 | 84.1 | 13.6 |
| PSM+MAP(XS) | | | | | | -0.014 | -2.9 | 0.045 | 91.8 | 48.7 |
| PSW+MAP(L) | 0.013 | 2.7 | 0.031 | 78.3 | 8.6 | 0.003 | 0.6 | 0.055 | 81.2 | 1.2 |
| PSW+MAP(M) | 0.017 | 3.4 | 0.026 | 82.1 | 16.3 | 0.000 | 0.0 | 0.053 | 81.3 | 2.2 |
| PSW+MAP(S) | -0.002 | -0.5 | 0.015 | 92.9 | 65.3 | -0.015 | -3.1 | 0.047 | 83.1 | 11.5 |
| PSW+MAP(XS) | | | | | | -0.039 | -7.8 | 0.035 | 93.3 | 63.5 |
| PSS+CL | -0.035 | -6.9 | 0.000 | 23.0 | -51.3 | -0.035 | -7.1 | 0.000 | 24.6 | -56.8 |
| PSS+PP | -0.035 | -7.0 | 0.037 | 98.4 | 168.7 | -0.035 | -7.0 | 0.044 | 98.4 | 153.8 |
| MM | -0.014 | -2.7 | 0.040 | 97.7 | 277.6 | -0.016 | -3.1 | 0.036 | 98.1 | 135.9 |
| | Model specification3: incorrect model omitting three confounders | | | | | | | | | |
| PSM | -0.061 | -12.2 | 0.092 | 89.1 | 115.4 | -0.070 | -13.9 | 0.111 | 91.3 | 106.3 |
| PSW | -0.061 | -12.2 | 0.065 | 92.9 | 129.7 | -0.062 | -12.3 | 0.067 | 95.7 | 121.1 |
| PSM+MAP(L) | 0.002 | 0.3 | 0.039 | 77.7 | 8.7 | 0.001 | 0.2 | 0.056 | 80.6 | 1.3 |
| PSM+MAP(M) | -0.001 | -0.3 | 0.041 | 79.7 | 15.4 | -0.003 | -0.5 | 0.054 | 80.8 | 2.2 |
| PSM+MAP(S) | -0.026 | -5.3 | 0.042 | 84.1 | 44.9 | -0.024 | -4.8 | 0.053 | 81.2 | 11.7 |
| PSM+MAP(XS) | | | | | | -0.042 | -8.5 | 0.069 | 84.9 | 39.2 |
| PSW+MAP(L) | 0.007 | 1.5 | 0.035 | 77.0 | 8.2 | 0.002 | 0.5 | 0.054 | 80.7 | 1.2 |
| PSW+MAP(M) | 0.007 | 1.3 | 0.030 | 80.9 | 15.7 | -0.001 | -0.2 | 0.055 | 80.5 | 2.0 |
| PSW+MAP(S) | -0.030 | -6.1 | 0.036 | 89.1 | 58.1 | -0.022 | -4.4 | 0.048 | 81.5 | 10.6 |
| PSW+MAP(XS) | | | | | | -0.060 | -12.0 | 0.068 | 88.7 | 49.4 |
| PSS+CL | -0.078 | -15.5 | 0.000 | 12.9 | -53.0 | -0.081 | -16.2 | 0.000 | 12.6 | -58.4 |
| PSS+PP | -0.078 | -15.6 | 0.088 | 93.2 | 150.5 | -0.080 | -16.1 | 0.089 | 95.4 | 137.1 |
| MM | -0.018 | -3.6 | 0.038 | 93.6 | 108.9 | -0.013 | -2.7 | 0.041 | 89.4 | 39.6 |

Unadj, unadjusted; rc, reduced concurrent trial (with half concurrent controls); fc, full concurrent trial (with full concurrent controls); ESSR, effective sample size rate; PSM, propensity score matching; PSW, propensity score weighting; PSS, propensity score stratification; PP, power prior; CL, composite likelihood; MAP, meta-analytic-predictive prior; MM, mixed effect model; MM.nc, mixed effect model with no covariates.

**Table 6** Summary of type 1 error and power for simulation with multiple historical trials

| | Type 1 error | | | | | | Power | | | | | |
|---|---|---|---|---|---|---|---|---|---|---|---|---|
| | Moderate | | | Severe | | | Moderate | | | Severe | | |
| | Model1 | Model2 | Model3 | Model1 | Model2 | Model3 | Model1 | Model2 | Model3 | Model1 | Model2 | Model3 |
| unadj.rc | 0.040 | | | 0.054 | | | 72.8 | | | 79.7 | | |
| unadj.fc | 0.053 | | | 0.048 | | | 88.5 | | | 91.4 | | |
| MAP(L) | 0.033 | | | 0.056 | | | 75.5 | | | 80.7 | | |
| MAP(M) | 0.034 | | | 0.054 | | | 78.2 | | | 80.6 | | |
| MAP(S) | 0.078 | | | 0.053 | | | 77.6 | | | 80.1 | | |
| MAP(XS) | | | | 0.125 | | | | | | 77.8 | | |
| MM.nc | 0.027 | | | 0.052 | | | 78.8 | | | 79.9 | | |
| MM | 0.043 | 0.040 | 0.038 | 0.046 | 0.036 | 0.041 | 98.8 | 97.7 | 93.6 | 99.6 | 98.1 | 89.4 |
| PSM | 0.045 | 0.057 | 0.092 | 0.042 | 0.048 | 0.111 | 97.7 | 95.6 | 89.1 | 99.3 | 97.7 | 91.3 |
| PSW | 0.015 | 0.025 | 0.065 | 0.028 | 0.028 | 0.067 | 98.1 | 96.8 | 92.9 | 99.6 | 98.8 | 95.7 |
| PSS+PP | 0.029 | 0.037 | 0.088 | 0.030 | 0.044 | 0.089 | 99.0 | 98.4 | 93.2 | 99.3 | 98.4 | 95.4 |
| PSM+MAP(L) | 0.035 | 0.031 | 0.039 | 0.054 | 0.053 | 0.056 | 77.3 | 77.8 | 77.7 | 80.8 | 81.0 | 80.6 |
| PSM+MAP(M) | 0.025 | 0.027 | 0.041 | 0.053 | 0.055 | 0.054 | 81.5 | 80.8 | 79.7 | 81.2 | 81.8 | 80.8 |
| PSM+MAP(S) | 0.026 | 0.031 | 0.042 | 0.042 | 0.047 | 0.053 | 90.0 | 88.9 | 84.1 | 84.3 | 84.1 | 81.2 |
| PSM+MAP(XS) | | | | 0.046 | 0.045 | 0.069 | | | | 91.8 | 91.8 | 84.9 |
| PSW+MAP(L) | 0.034 | 0.031 | 0.035 | 0.056 | 0.055 | 0.054 | 78.5 | 78.3 | 77.0 | 81.2 | 81.2 | 80.7 |
| PSW+MAP(M) | 0.028 | 0.026 | 0.030 | 0.054 | 0.053 | 0.055 | 83.1 | 82.1 | 80.9 | 81.4 | 81.3 | 80.5 |
| PSW+MAP(S) | 0.009 | 0.015 | 0.036 | 0.045 | 0.047 | 0.048 | 93.9 | 92.9 | 89.1 | 83.8 | 83.1 | 81.5 |
| PSW+MAP(XS) | | | | 0.027 | 0.035 | 0.068 | | | | 95.0 | 93.3 | 88.7 |

The colours represent the magnitude of Type I error (ranging from light yellow to red in the left panel) and power (ranging from white to green in the right panel) from low to high. The results are shown for two scenarios, simulating moderate and severe among-study heterogeneities, respectively. Model1: the correct model specifications; Model2: the incorrect model specification, achieved by removing one confounder when confounders are required; Model3 the incorrect model specification, achieved by removing three confounders when confounders are required. Unadj, unadjusted; rc, reduced concurrent trial (with half concurrent controls); fc, full concurrent trial (with full concurrent controls); ESSR, effective sample size rate; PSM, propensity score matching; PSW, propensity score weighting; PSS, propensity score stratification; PP, power prior; MAP, meta-analytic-predictive prior; MM, mixed effect model; MM.nc, mixed effect model with no covariates.

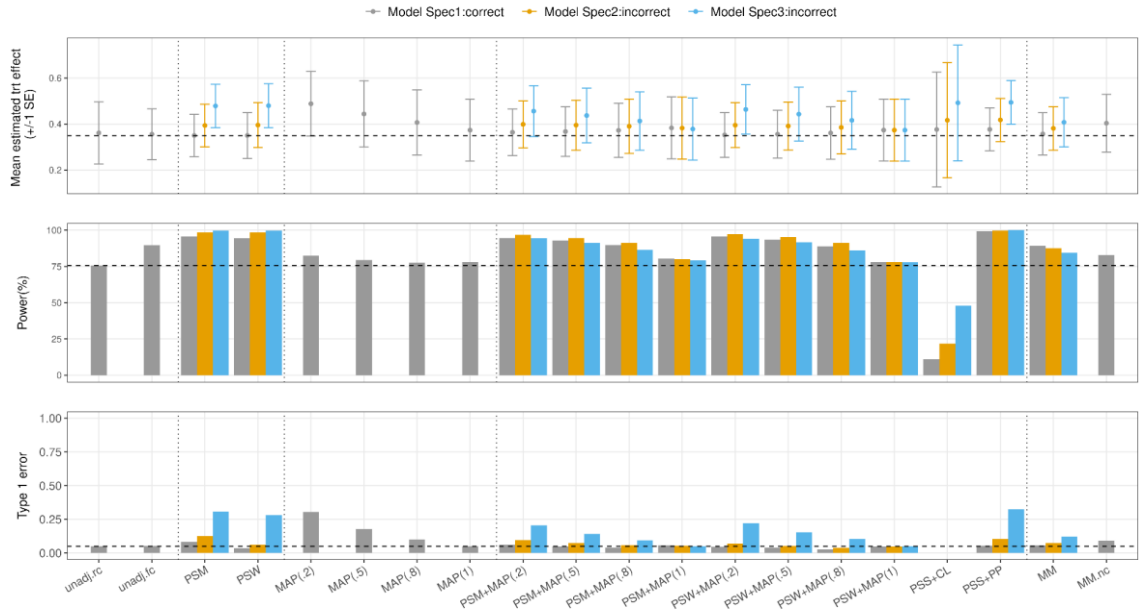

**Fig 3** Simulation results demonstrate scenarios involving a single historical trials and moderate heterogeneities between the concurrent and historical trials. The upper, middle, and bottom panels display the mean estimated treatment effect along with the corresponding ± 1 estimated standard error, power under the alternative hypothesis, and Type 1 error under the null hypothesis, respectively. Dashed horizontal lines represent the true effect size ($\theta_{trt} = 0.5$), power for the unadjusted model with concurrent treated and control data (approximately 75%), and Type 1 error at a level of 0.5. The results of different model specifications are shown in different colors; (grey) unadjusted model without any confounders or correct model specifications with all six confounders, (orange) incorrect model specification 1 which includes five confounders and omits confounder $x_4$, and (blue) incorrect model specification 2 which includes three confounders ($x_1, x_2, x_3$). Unadj, unadjusted; rc, reduced concurrent trial (with half concurrent controls); fc, full concurrent trial (with full concurrent controls); PSM, propensity score matching; PSW, propensity score weighting; PSS, propensity score stratification; PP, power prior; CL, composite likelihood; MAP, meta-analytic-predictive prior; MM, mixed effect model; MM.nc, mixed effect model with no covariates.

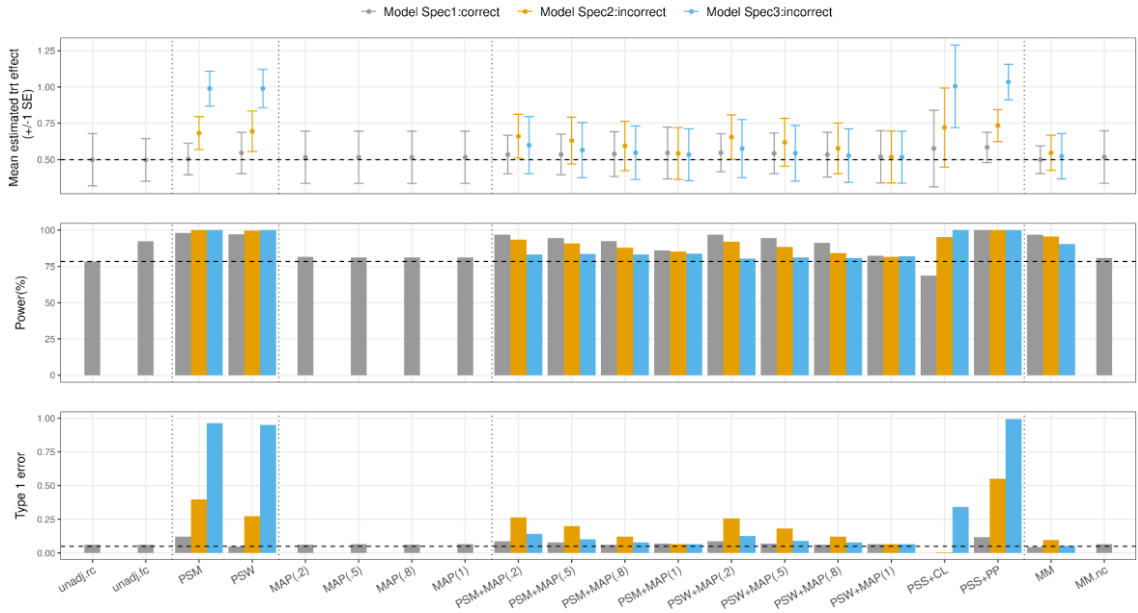

**Fig 4** Simulation results demonstrate scenarios involving a single historical trials and severe heterogeneities between the concurrent and historical trials. The upper, middle, and bottom panels display the mean estimated treatment effect along with the corresponding ± 1 estimated standard error, power under the alternative hypothesis, and Type 1 error under the null hypothesis, respectively. Dashed horizontal lines represent the true effect size ($\theta_{trt} = 0.5$), power for the unadjusted model with concurrent treated and control data (approximately 75%), and Type 1 error at a level of 0.5. The results of different model specifications are shown in different colors; (grey) unadjusted model without any confounders or correct model specifications with all six confounders, (orange) incorrect model specification 1 which includes five confounders and omits confounder $x_4$, and (blue) incorrect model specification 2 which includes three confounders ($x_1, x_2, x_3$). Unadj, unadjusted; rc, reduced concurrent trial (with half concurrent controls); fc, full concurrent trial (with full concurrent controls); PSM, propensity score matching; PSW, propensity score weighting; PSS, propensity score stratification; PP, power prior; CL, composite likelihood; MAP, meta-analytic-predictive prior; MM, mixed effect model; MM.nc, mixed effect model with no covariates.

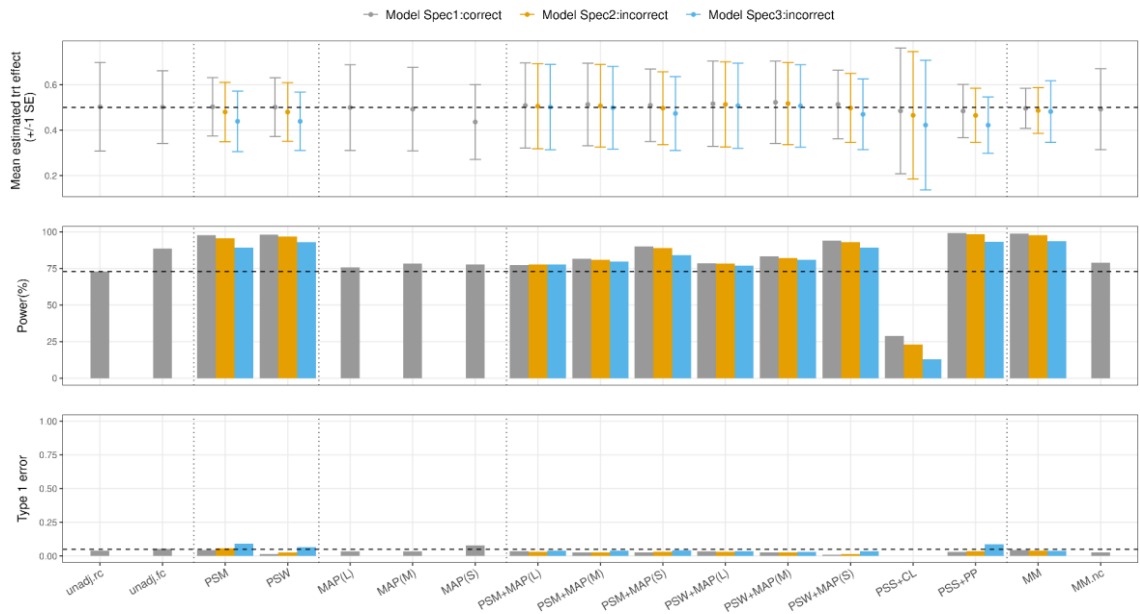

**Fig 5** Simulation results demonstrate scenarios involving multiple historical trials and moderate heterogeneities between the concurrent and historical trials. The upper, middle, and bottom panels display the mean estimated treatment effect along with the corresponding ± 1 estimated standard error, power under the alternative hypothesis, and Type 1 error under the null hypothesis, respectively. Dashed horizontal lines represent the true effect size ($\theta_{trt} = 0.5$), power for the unadjusted model with concurrent treated and control data (approximately 75%), and Type 1 error at a level of 0.5. The results of different model specifications are shown in different colors; (grey) unadjusted model without any confounders or correct model specifications with all six confounders, (orange) incorrect model specification 1 which includes five confounders and omits confounder $x_4$, and (blue) incorrect model specification 2 which includes three confounders ($x_1, x_2, x_3$). Unadj, unadjusted; rc, reduced concurrent trial (with half concurrent controls); fc, full concurrent trial (with full concurrent controls); PSM, propensity score matching; PSW, propensity score weighting; PSS, propensity score stratification; PP, power prior; CL, composite likelihood; MAP, meta-analytic-predictive prior; L, M, S, XS, large, medium, small and extra small weight for MAP, MM, mixed effect model; MM.nc, mixed effect model with no covariates.

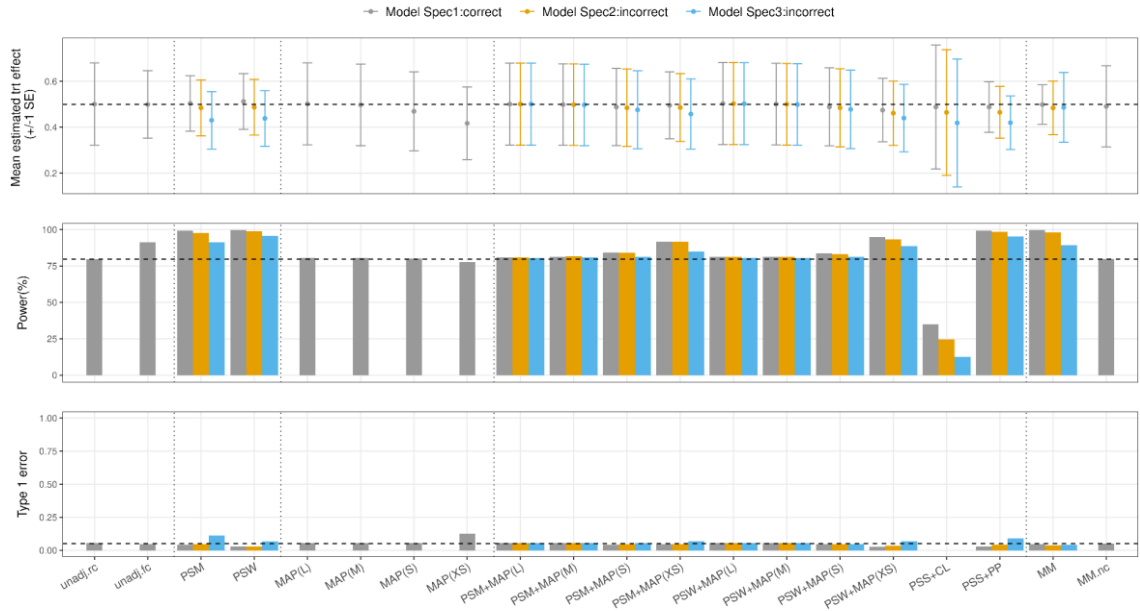

**Fig 6** Simulation results demonstrate scenarios involving multiple historical trials and severe heterogeneities between the concurrent and historical trials. PSS+CL is included in this plot. The upper, middle, and bottom panels display the mean estimated treatment effect along with the corresponding ± 1 estimated standard error, power under the alternative hypothesis, and Type 1 error under the null hypothesis, respectively. Dashed horizontal lines represent the true effect size ($\theta_{trt} = 0.5$), power for the unadjusted model with concurrent treated and control data (approximately 75%), and Type 1 error at a level of 0.5. The results of different model specifications are shown in different colors; (grey) unadjusted model without any confounders or correct model specifications with all six confounders, (orange) incorrect model specification 1 which includes five confounders and omits confounder $x_4$, and (blue) incorrect model specification 2 which includes three confounders ($x_1, x_2, x_3$). Unadj, unadjusted; rc, reduced concurrent trial (with half concurrent controls); fc, full concurrent trial (with full concurrent controls); PSM, propensity score matching; PSW, propensity score weighting; PSS, propensity score stratification; PP, power prior; CL, composite likelihood; MAP, meta-analytic-predictive prior; L, M, S, XS, large, medium, small and extra small weight for MAP, MM, mixed effect model; MM.nc, mixed effect model with no covariates.